\documentclass{article}
\usepackage{spconf,amsmath,graphicx,hyperref,booktabs,multirow,xcolor,array,arydshln}

\usepackage{adjustbox}
\usepackage{array}
\usepackage{tabularx}
\usepackage{minted}
\usepackage[most]{tcolorbox}
\usepackage{enumitem}
\usepackage{multicol}
\usepackage{listings}
\usepackage{float}

\title{FoleyBench: A Benchmark For Video-to-Audio Models}
\name{Satvik Dixit$^{1}$, Koichi Saito$^{2}$, Zhi Zhong$^{3}$, Yuki Mitsufuji$^{2,3}$, Chris Donahue$^{1}$}
  
\address{$^{1}$Carnegie Mellon University, 
         $^{2}$Sony AI, $^{3}$Sony Group Corporation }
    
\begin{document}
\ninept

\maketitle

\begin{abstract}
\vspace{-0.05 in}
Video-to-audio generation (V2A) is of increasing importance in domains such as film post-production, AR/VR, and sound design, particularly for the creation of \emph{Foley}---sound effects synchronized with on-screen actions. Foley requires generating audio that is both semantically aligned with visible events and temporally aligned with their timing. Yet, there is a mismatch between evaluation and downstream applications due to the absence of a benchmark tailored to Foley-style scenarios. We find that 74\% of videos from past evaluation datasets have poor audio-visual correspondence. Moreover, they are dominated by speech and music—domains that lie outside the use case for Foley.
To address this gap, we introduce FoleyBench, the first large-scale benchmark explicitly designed for Foley-style V2A evaluation. FoleyBench contains 5,000 (video, ground-truth audio, text caption) triplets, each featuring visible sound sources with audio causally tied to on-screen events. The dataset is built using an automated, scalable pipeline applied to in-the-wild internet videos from YouTube-based and Vimeo-based sources. Compared to past datasets, we show that videos from FoleyBench have stronger coverage of sound categories from a taxonomy specifically designed for Foley sound. Each clip is further labeled with metadata capturing source complexity, UCS/AudioSet category, and video length, enabling fine-grained analysis of model performance and failure modes. We benchmark state-of-the-art (SotA) V2A models, evaluating them on audio quality, audio–video alignment, temporal synchronization, and audio–text consistency. Samples are available at: \url{https://gclef-cmu.org/foleybench}.

\end{abstract}
\begin{keywords}
Video-to-audio models, Audio-visual alignment, Foley sound synthesis
\end{keywords}
\vspace{-0.1 in}
\section{Introduction}
\label{introduction}
\vspace{-0.1 in}
\begin{figure*}
\centering
\includegraphics[width=1\linewidth]{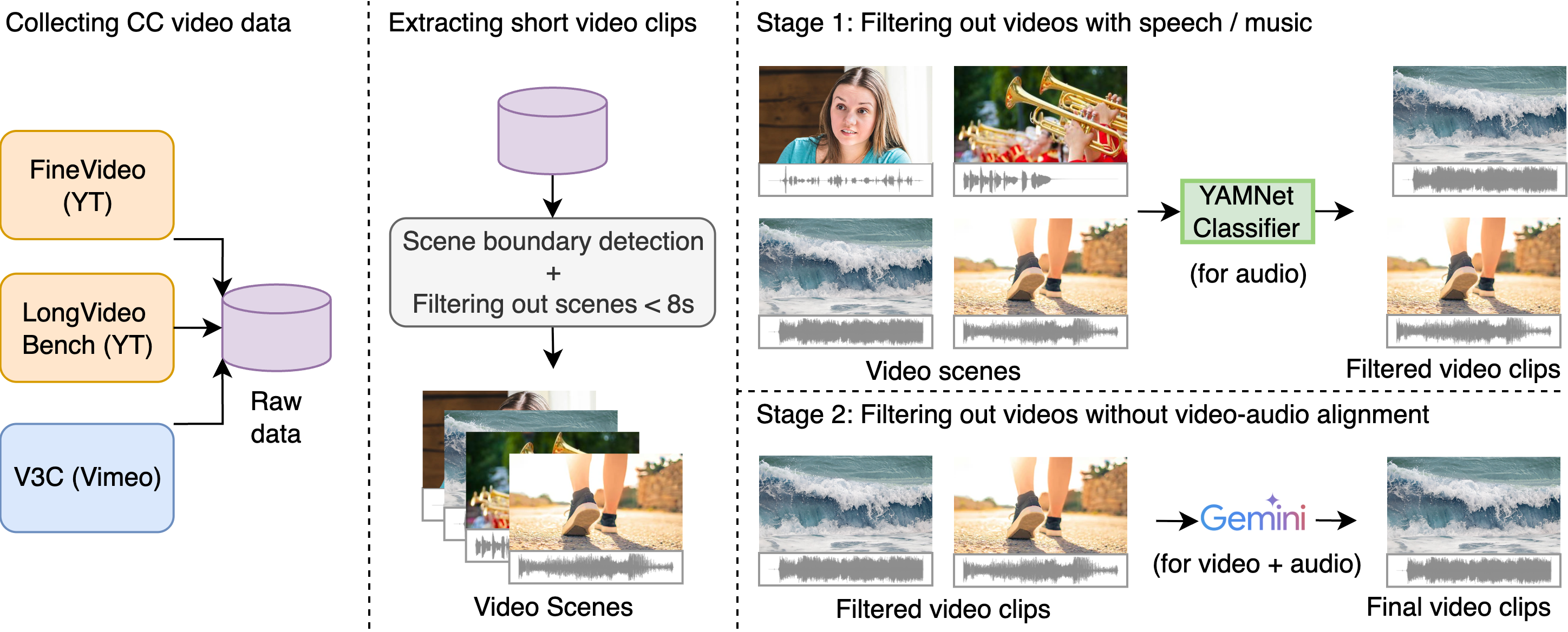}
\vspace{-5pt}
\caption{\textbf{Dataset construction pipeline.} Raw videos are first collected from Creative Commons (CC) licensed sources.
Scenes are detected, trimmed to start/end timestamps, and discarded if shorter than 8s. Content-based filtering is then done in two stages: (1) YAMNet removes clips with speech or music, and (2) Gemini discards clips where audio is not visually or causally grounded, yielding high-quality Foley clips.}
\vspace{-5pt}
\label{fig:placeholder}
\end{figure*}

The task of video-to-audio generation (V2A) seeks to synthesize realistic, temporally aligned audio directly from visual input, sometimes conditioned on an additional text caption for finer control. A key downstream use case in the creative industry (e.g., film, games) is Foley—the creation of sound effects such as object impacts, or ambient noises that must be both semantically appropriate and temporally synchronized with visible events. Foley is distinct from speech and music, which are typically added in other stages of production. Foley specifically refers to non-speech, non-music sound effects (like footsteps or object impacts) that are causally linked to visual actions. Recent work has rapidly advanced V2A generation to support Foley scenarios, with a diverse family of methods being developed, including autoregressive, diffusion-based, flow-matching-based, and ControlNet-based models \cite{Mei2023-FoleyGen, viertola2025temporally, luo2023diff, xu2024video, benita2025cafa, cheng2025mmaudio, chen2025video}.

Despite this progress, evaluation of V2A systems remains misaligned with downstream Foley application goals. The widely used VGGSound test set~\cite{vggsound2020}, for instance, suffers from weak audio–visual correspondence, with many clips dominated by speech, music, or off-screen sounds. In our analysis, we find that 74\% of these clips show poor alignment between the audio and the video, as detailed in Section ~\ref{subsec:comparison}.
These cases lie outside the intended scope of V2A modeling, and their prevalence in existing benchmarks obscures evaluation.
Other relevant datasets exist but are small-scale, limited in diversity, or lack comprehensive metadata.
Three existing datasets~\cite{chen2020generating,iashin2022sparse,owens2016visually} have stronger guarantees on audio-visual correspondence but only cover a dozen or fewer specific sound categories.
The recently proposed VisualSound~\cite{viertola2025temporally} dataset filters VGGSound using an audio-visual similarity metric,
but lacks comprehensive sound category metadata, and its coverage of Foley categories is unknown.

Furthermore, current benchmarks lack systematic coverage across sound classes (such as AudioSet~\cite{audioset2017} or the Universal Category System (UCS)~\cite{UCS}—a sound categorization developed specifically for professional sound design use cases).
They lack categories for fine-grained evaluation, such as sound types (discrete events vs. continuous ambiences) and scene complexity (single-source vs. multi-source), limiting our understanding of where models succeed or fail.
Our experiments on the VGGSound test set reveal its coverage across Foley sound categories is limited—24.3\% of UCS categories have three or fewer videos in VGGSound.

To address these challenges, we introduce FoleyBench, the first large-scale benchmark explicitly designed for evaluating Foley-style V2A: non-speech, non-music clips with visible sound sources and strong temporal correspondence between audio and video. FoleyBench comprises 5,000 curated video–audio–caption triplets, each approximately 8 to 10 seconds long, constructed through a scalable multi-stage pipeline. Each clip is automatically labeled by Gemini 2.5 Pro \cite{GoogleDeepMind2025Gemini2_5} with AudioSet \cite{audioset2017} and UCS tags, and enriched with metadata capturing source complexity (such as single vs. multi-source) and sound envelope type (such as discrete vs. continuous). We also create FoleyBench-Long—a set of 650 long videos (each 30 seconds long) to evaluate the audio generation ability of models for long-form videos. All these categories enable fine-grained analysis of model behavior across conditions. 
We find that while MMAudio \cite{cheng2025mmaudio} achieves the best overall scores, other models excel in specific areas: Seeing \& Hearing \cite{xing2024seeing} obtains the best semantic audio-video alignment, V-AURA \cite{viertola2025temporally} excels at precise temporal synchronization, and LOVA \cite{cheng2025lova} better preserves audio quality on long-form content. Our contributions are as follows:
\vspace{-0.05 in}
\begin{itemize}[noitemsep]
\item FoleyBench, a 5,000-instance, high-quality benchmark for visually grounded, non-speech, non-music audio generation.
\item An automatic, multi-stage pipeline for filtering Foley-style clips from in-the-wild videos.
\item Comprehensive evaluation of V2A models on audio quality, audio–visual alignment, temporal synchronization, and audio–text consistency.
\item Analyzing failure modes in existing V2A models, yielding actionable insights for future progress.
\end{itemize}

\vspace{-0.1 in}
\section{FoleyBench}
\vspace{-0.1 in}
FoleyBench is a large-scale benchmark designed to evaluate V2A systems in scenarios comprising video clips with precise temporal and semantic grounding. The dataset focuses on \emph{Foley-style clips}, which we define here as: non-speech, non-music video segments in which the sound sources are both visible on screen and temporally synchronized with their corresponding actions.
It consists of 5,000 curated video–audio pairs, each accompanied by a caption describing the visual scene. Importantly, each clip is accompanied by metadata that enables fine-grained analysis of model behavior. Metadata includes: AudioSet and UCS labels and attributes such as sound type (discrete events vs. continuous ambience) and source complexity (single-source vs. multi-source).

To construct FoleyBench, we developed an automated
multi-stage pipeline for filtering Foley-style clips from large collections of internet videos as shown in Figure~\ref{fig:placeholder}. The dataset is constructed through the following stages:
\begin{enumerate}[leftmargin=*]%

    \item \textbf{Data Collection:} Raw videos are gathered from diverse, internet-scale, Creative Commons (CC) licensed sources to cover a diverse range of scenarios: YouTube (FineVideo \cite{finevideo2024}, LVBench \cite{wang2024lvbench}) and Vimeo (V3C1 \cite{berns2019v3c1}).
    \item \textbf{Scene Detection:} Each video is segmented into scenes using automatic scene boundary detection. Scenes shorter than 8s are discarded to avoid overly short contexts.
    \item \textbf{Content-based Filtering:} To ensure non-speech, non-music clips with strong audio-visual correspondence, we apply a two-stage filtering pipeline:
        \begin{enumerate}[leftmargin=*]%
            \item \textbf{Audio filtering:} We first apply YAMNet \cite{yamnet2020} for frame-level classification. We use YAMNet as a fast and cost-effective method for a coarse-grained filtering of the majority of irrelevant clips. 
            Any clip containing speech or music label score over 0.6 at any frame is discarded. This removes unwanted cases, even when speech/music is present for a brief interval or only present in the background. Since in-the-wild videos are dominated by speech and music, this step filters the vast majority (97.7\%) of clips.  We manually annotate 276 videos sampled uniformly at random from the output of this filtering and find that 130 of them were Foley-style clips, i.e.~the precision of this filtering in terms of our downstream goal is~47\%. Most false positives were for non-speech non-music videos where the audio was not causally linked to the on screen events.
            \item \textbf{Audiovisual filtering:}  Remaining 
            clips are evaluated using Gemini 2.5 Pro \cite{GoogleDeepMind2025Gemini2_5}, which judges whether the sounds are causally and temporally grounded in visible on-screen actions. For example, if the audio is clapping, the video must show visible hands clapping in sync. We pass our manually annotated set of 276 clips through this stage and observe 72\% precision. This stage substantially improves the quality of the dataset by filtering out mismatched or weakly grounded cases. 
        \end{enumerate}
\end{enumerate}
\vspace{-0.1 in}
\section{Experimental Setup}
\vspace{-0.1 in}
\begin{table*}[!t]
\caption{{\bf Evaluation of V2A models on VGGSound test set and FoleyBench.}
We report results across six metrics grouped into cross-modal alignment (ImageBind (IB), CLAP, De-Sync (DS)) and audio quality (FAD, IS, KLD). IB and CLAP scores are between 0 and 1, and De-Sync is measured in seconds. CLAP is only measured for models that condition on the text prompt (marked with superscript T). * Denotes evaluation on a filtered subset of the VGGSound test set. The best results for each metric are shown in bold and second-best are underlined.}

\label{tab:video2audio}

\centering
\footnotesize
\setlength\tabcolsep{7pt}
\renewcommand{\arraystretch}{1.0}
\newcommand\padd{\phantom{0}}

\resizebox{1\textwidth}{!}{

\begin{tabular}{l|ccc ccc|ccc ccc}
\toprule
\multirow{3}{*}{Method} 
& \multicolumn{6}{c|}{VGGSound test set} 
& \multicolumn{6}{c}{FoleyBench} \\
\cmidrule(lr){2-7} \cmidrule(lr){8-13}
& \multicolumn{3}{c}{Cross-modal alignment} & \multicolumn{3}{c|}{Audio quality} 
& \multicolumn{3}{c}{Cross-modal alignment} & \multicolumn{3}{c}{Audio quality} \\
\cmidrule(lr){2-4} \cmidrule(lr){5-7} \cmidrule(lr){8-10} \cmidrule(lr){11-13} 
& IB~$\uparrow$ & CLAP~$\uparrow$ & DS~$\downarrow$ 
& FAD~$\downarrow$ & IS~$\uparrow$ & KLD~$\downarrow$ 
& IB~$\uparrow$ & CLAP~$\uparrow$ & DS~$\downarrow$ 
& FAD~$\downarrow$ & IS~$\uparrow$ & KLD~$\downarrow$ \\
\midrule
V-AURA~\cite{viertola2025temporally}& 0.276& -- & 0.654& 14.80& 10.08& 2.42& 0.237 & --& \underline{0.716} & 27.2 & 6.44 & 3.46 \\
% solid, 695M, no text
DiffFoley~\cite{luo2023diff}      & -- & -- & -- & -- & -- & -- & 0.173 & --& 0.88 & 31.9 & 9.26 & 3.88 \\
% only have results on clean VGG, size?,no text
Seeing\&Hearing~\cite{xing2024seeing} & \textbf{0.339} & -- & 1.204& 24.58& 8.58& 2.26& \textbf{0.371} & --& 1.08 & 25.0& 4.80& 3.30 \\
% solid, 415M, no text
MaskVAT~\cite{pascual2024masked} & -- & -- & -- & -- & -- & -- & 0.239 & --& 0.748 & 19.7 & 6.94 & 3.22 \\
% no numbers, size?, no text
V2A-Mapper~\cite{wang2024v2a} & 0.226 & -- & 1.225 & 8.40 & 12.47 & 2.69 & 0.189 & --& 1.09 & 16.2 & 8.87 & 3.50 \\
% solid, 229M, no text
SpecMaskFoley\textsuperscript{T}~\cite{zhong2025specmaskfoley}      & 0.264 & -- & \underline{0.652} & 7.425 & 11.43& 1.98 & 0.229 & 0.191 & 0.801 & 19.2 & 5.86 & 3.08 \\
% solid, 300M, text
VTA-LDM\textsuperscript{T}~\cite{xu2024video}      & -- & \textbf{0.452}& -- & \textbf{2.05} & 10.10& 3.81& 0.221 & 0.138 & 1.21 & 15.7 & 7.27 & 3.13 \\
FoleyCrafter\textsuperscript{T}~\cite{zhang2024foleycrafter}      & 0.257 & -- & 1.225 & 16.24 & \underline{15.68} & 2.30 & 0.255 & 0.261& 1.15 & 16.5 & \underline{9.50} & 2.68 \\
% solid, 1.22B, text
LOVA\textsuperscript{T}~\cite{cheng2025lova}      & -- & -- & -- & -- & 9.73 & 2.06 & 0.209 & 0.167 & 1.15 & 20.7 & 7.61 & 3.15 \\
% some available, size?, text
CAFA\textsuperscript{T}~\cite{benita2025cafa}      &  0.210 & 0.230 & 0.960 & 12.60 & 13.45 & 2.02 & 0.198 & \underline{0.270} & 0.825 & \underline{15.5} & 7.41 & \underline{2.54} \\
MultiFoley\textsuperscript{T}~\cite{chen2025video} & 0.280\textsuperscript{*} & \underline{0.344\textsuperscript{*}} & -- & -- & -- & \textbf{1.43\textsuperscript{*}} & 0.144 & 0.218 & 0.846 & 22.3 & 5.38 & 2.75 \\
% solid, size?, text
MMAudio\textsuperscript{T}~\cite{cheng2025mmaudio}      & \underline{0.332} & -- & \textbf{0.442} & \underline{4.72} &\textbf{ 17.40}& \underline{1.65} & \underline{0.306} & \textbf{0.331} & \textbf{0.447} & \textbf{8.76} & \textbf{11.2} & \textbf{2.43} \\
% solid, 1.03B, text
\bottomrule
\end{tabular}
}
\vspace{-0.1in}
\end{table*}

We benchmark FoleyBench against a diverse set of recent V2A generation models, across a range of architectures. Our evaluation includes an autoregressive model (V-AURA~\cite{viertola2025temporally}), which predicts audio tokens sequentially; several diffusion-based approaches (DiffFoley~\cite{luo2023diff}, VTA-LDM~\cite{xu2024video}, FoleyCrafter~\cite{zhang2024foleycrafter}, and LOVA~\cite{cheng2025lova}), which iteratively refine audio from noise and often emphasize temporal alignment or long-form synthesis; and adapter-based methods such as CAFA~\cite{benita2025cafa}, which guide a pretrained text-to-audio generator through additional control modules. We also consider transformer-based masked prediction models (MaskVAT~\cite{pascual2024masked}, SpecMaskFoley~\cite{zhong2025specmaskfoley}, and Seeing\&Hearing~\cite{xing2024seeing}), which adapt the MaskGIT framework to audio by filling in masked spectrogram regions and in some cases leverage test-time optimization for better grounding. We also have V2A-Mapper \cite{wang2024v2a} that uses embedding translation, which projects CLIP video embeddings into the CLAP audio–text space to condition an AudioLDM backbone without extensive retraining. 
Finally, we evaluate two large-scale, jointly-trained models: MultiFoley~\cite{chen2025video}, a diffusion transformer trained on internet videos and high-quality SFX libraries, and MMAudio~\cite{cheng2025mmaudio}, a large flow-matching model that jointly trains across video, text, and audio modalities. Together, these baselines reflect the breadth of strategies explored in V2A research—from lightweight adapters to fully joint multimodal models.

\noindent \textbf{Evaluation Metrics.}
We evaluate generated audio along two key dimensions: audio quality and cross-modal alignment. For audio quality, we report three widely used metrics: Fréchet Audio Distance (FAD)~\cite{fad2019}, Inception Score (IS)~\cite{salimans2016inception}, and Kullback–Leibler Divergence (KLD)~\cite{Kullback1951KL}. To compute all of these metrics, we rely on PANN embeddings~\cite{kong2020panns}, which provide pretrained audio representations especially effective for non-speech, non-music audio. For alignment, we focus on three complementary metrics. CLAP Score~\cite{laionclap2023} evaluates how well generated audio semantically matches the ground-truth textual captions. As this metric specifically measures the generated audio's adherence to the text prompt, we only report it for models that are text-conditioned (marked with \textsuperscript{T} in Table~\ref{tab:video2audio}). ImageBind Score~\cite{girdhar2023imagebind} measures cross-modal similarity between audio and video. Finally, De-Sync~\cite{syncnet2016} quantifies the temporal offset between audio and video streams in seconds, with lower scores corresponding to tighter synchronization. We use the AV-benchmark toolkit for evaluation \footnote{\url{https://github.com/hkchengrex/av-benchmark}}.

\vspace{-0.2 in}
\section{Results}
\label{results}
\vspace{-0.1 in}
We evaluate FoleyBench along two axes: (1) its reliability as a Foley benchmark compared to existing datasets in Section \ref{subsec:comparison}, and (2) the insights it provides into the behavior of V2A models in Section \ref{subsec:evaluation}.
\vspace{-0.1 in}
\subsection{Comparison with VGGSound-Test} \label{subsec:comparison}
\vspace{-0.05 in}
We compare FoleyBench to the VGGSound test set, the \emph{de facto}
standard for evaluating V2A models, to demonstrate that it is ill-suited for Foley-style scenarios and that FoleyBench is a more reliable alternative.

\noindent \textbf{Dataset quality.}  A key limitation of VGGSound-Test is the prevalence of unsuitable content for Foley evaluation. To assess this, we applied FoleyBench’s filtering pipeline to the VGGSound test set. We find that a significant portion of clips contain speech or music, with 65.3\% of the available videos being discarded by our initial audio filtering stage. After the subsequent Gemini-based audiovisual grounding filter, only 25.5\% of the original available videos remain (by comparison, we estimate that 72\% of the videos in our benchmark are relevant). This underscores that the majority of VGGSound is not representative of Foley-style scenarios. Another issue is availability-at present we find that only 84.3\% of the original VGGSound test set remains available on Youtube. Moreover, the distribution of sound categories in VGGSound-Test is less diverse. We pass the subset of VGGSound-Test remaining after the first stage of filtration through Gemini to get the UCS category label for each clip. Using Gemini to assign UCS category labels, we find that 24.3\% of categories contain three or fewer videos, compared to only 13.4\% in FoleyBench. We measure the overall diversity of the distribution of VGGSound-Test over UCS categories by calculating the Shannon entropy of the distribution across categories (higher is more diverse). The Shannon entropy of the filtered subset of VGGSound-Test is 4.73 compared to 5.35 for FoleyBench. 
This lack of suitable, diverse content makes VGGSound-test ill-suited for V2A model evaluation.

\noindent \textbf{Correlation of metric scores.}  
We evaluate SotA V2A models on both FoleyBench and VGGSound test using 6 metrics: CLAP Score (audio-text similarity), ImageBind Score (audio-video similarity), and De-Sync (temporal synchronization), and three audio fidelity metrics—FAD, Inception Score (IS), and KL Divergence. For each metric, we compute Kendall’s rank correlation ($\rho$) between the scores of each model on FoleyBench and VGGSound Test.
We find that correlation is relatively strong for De-Sync ($\rho$ = 0.878, p = 0.006) and ImageBind Score ($\rho$ = 0.714, p = 0.030), suggesting some consistency in measuring temporal synchronization and audio-video alignment. Correlations are lower for audio fidelity metrics: FAD ($\rho$ = 0.429, p = 0.179), Inception Score ($\rho$ = 0.611, p = 0.025), and KL Divergence ($\rho$ = 0.556, p = 0.045). We do not compute the correlation for CLAP score as there is not enough data for this metric on VGGSound Test. These results highlight that FoleyBench yields meaningfully different evaluation outcomes compared to VGGSound, particularly on audio fidelity metrics where correlations are weaker.  
We hypothesize audio fidelity metrics on our dataset may be more sensitive to performance on broad sound classes, since we have higher coverage of UCS categories.

\vspace{-0.1 in}
\subsection{New Insights from FoleyBench} \label{subsec:evaluation}
\vspace{-0.05 in}
We present the benchmark results for several SotA V2A models on FoleyBench in Table~\ref{tab:video2audio}, providing a comprehensive set of reference evaluations. These scores are shown alongside those from the VGGSound Test set, which are often sparsely and inconsistently reported in prior work—a key gap our benchmark aims to resolve.
For FoleyBench, we find that MMAudio \cite{cheng2025mmaudio} achieves the best performance across nearly all metrics (second-best only on IB). Seeing and Hearing~\cite{xing2024seeing} achieves the best ImageBind score, which is likely because it performs gradient ascent directly on the ImageBind score at test-time, as also noted by \cite{cheng2025mmaudio}. Interestingly, CAFA \cite{benita2025cafa}, a ControlNet-based approach, ranks second across multiple metrics. This highlights the effectiveness of adapting a strong pretrained backbone (Stable-Audio-Open \cite{evans2025stable}), which provides a full-band (44.1kHz) output, offering an advantage in audio quality metrics over models with lower-bandwidth backbones (e.g., SpecMaskGIT at 22kHz). V-AURA \cite{viertola2025temporally} ranks second on temporal synchronization (DS), consistent with its design emphasis on fine-grained temporal alignment via high-frame-rate visual feature extraction. 

Crucially, FoleyBench’s metadata enables a more fine-grained analysis of model behavior beyond these aggregate scores, revealing several insights. In the following analysis, we focus primarily on MMAudio, as it achieved the strongest overall performance.

\noindent \textbf{Discrete Events.} We find that models know \emph{when} a discrete sound happens, but not \emph{what} it should sound like. On \texttt{Discrete} clips (e.g., impacts, snaps) compared to \texttt{Rest} (e.g., continuous sound like wind, rain), models show a trade-off: temporal synchronization improves, but audio quality deteriorates. For example, MMAudio's De-Sync improves (0.458s $\to$ 0.390s), but its FAD degrades severely (9.02 $\to$ 16.35), while its IS also drops (10.3 $\to$ 8.8). We observe this same pattern in other models as well: CAFA~\cite{benita2025cafa} shows De-Sync improving (0.842s $\to$ 0.730s) while FAD degrades (16.06 $\to$ 22.15); V-AURA~\cite{viertola2025temporally} shows De-Sync improving (0.812s $\to$ 0.736s) as FAD also degrades (59.17 $\to$ 64.39). This suggests that while the visual cue for an event provides a clear temporal signal, current models fail to render the corresponding high-fidelity impact.

\noindent \textbf{Background Sounds.} Models perform poorly on \texttt{Background} sounds. When comparing \texttt{Background} to \texttt{Action} clips, nearly all of MMAudio's metrics degrade: FAD worsens (9.77 $\to$ 14.76), IS drops (11.98 $\to$ 4.61), and De-Sync worsens (0.405s $\to$ 0.636s). However, the KLD (measuring class distribution) paradoxically \emph{improves} (2.54 $\to$ 1.98). This suggests that MMAudio can successfully identify the \emph{general category} of an ambient sound (e.g., "wind") but is currently unable to render it with high fidelity or synchronize it with subtle visual cues.

\noindent \textbf{Multisource Sounds.} As expected, \texttt{Multi-source} scenes are challenging. For CAFA, moving from \texttt{Single-source} to \texttt{Multi-source} degrades audio quality (FAD: 16.55 $\to$ 18.25), sync (De-Sync: 0.806s $\to$ 0.856s), and alignment (IB: 0.202 $\to$ 0.191). Interestingly, MMAudio exhibits a more nuanced failure: while its signal quality and sync also degrade (FAD: 9.84 $\to$ 11.34; De-Sync: 0.436s $\to$ 0.467s), its semantic metrics \emph{improve} (IB: 0.296 $\to$ 0.324). This suggests the model's low-quality audio "mashup" is a better semantic match for the combined visual concepts than a single, low-fidelity sound was.

\noindent \textbf{Text-Conditioning.} The results in Table \ref{tab:video2audio} suggest that text-conditioning acts as a powerful prior, as best performing models are text conditioned. We also performed an ablation on SpecMaskFoley~\cite{zhong2025specmaskfoley}, evaluating it with and without its text prompt. The text-conditioned version was universally superior across every metric. On \texttt{Action} clips, for instance, it improves audio quality metrics (FAD: 23.18 $\to$ 19.60; IS: 4.82 $\to$ 6.02) and alignment metrics (IB: 0.188 $\to$ 0.222; De-Sync: 0.911s $\to$ 0.755s). This confirms that text prompts provide crucial guidance, helping the model select a more accurate sound and align it more precisely in time.

\noindent \textbf{Long videos.} To assess the ability of V2A models to handle long-form generation, we introduce FoleyBench-Long, a collection of 650 videos each 30 seconds in length. This is a challenging task, as most V2A models are trained on clips of 10 seconds or less. Of the models we tested, only three—LOVA~\cite{cheng2025lova}, VTA-LDM~\cite{xu2024video}, and MMAudio~\cite{cheng2025mmaudio}—could produce 30-second outputs. As shown in Table \ref{tab:video2audiolong}, results degrade in general compared to short videos. MMAudio remains the best at synchronization (De-Sync) and semantic alignment (IB, CLAP), but its audio quality deteriorates substantially (FAD: 8.76 $\to$ 27.5). Conversely, LOVA, which was designed for long-form synthesis, achieves the best audio quality (FAD, IS) but at the cost of weaker alignment. These results show that generating temporally consistent and high-quality audio at 30-second scales remains a significant challenge. By providing FoleyBench-Long, we aim to establish a reliable benchmark for this underexplored setting and encourage future work on long-form V2A generation.

\begin{table}[!t]
\caption{\textbf{Evaluation of V2A models on FoleyBench-Long.} We report results on FoleyBench-Long and FoleyBench for models capable of generating long audios. Best results shown in bold.}
\label{tab:video2audiolong}

\centering
\footnotesize
\setlength\tabcolsep{7pt}
\renewcommand{\arraystretch}{1.0}
\newcommand\padd{\phantom{0}}

\resizebox{0.48\textwidth}{!}{

\begin{tabular}{l ccc ccc}
\toprule
\multirow{2}{*}{Method} 
& \multicolumn{3}{c}{Cross-modal alignment} & \multicolumn{3}{c}{Audio quality} \\
\cmidrule(lr){2-4} \cmidrule(lr){5-7} 
& IB~$\uparrow$ & CLAP~$\uparrow$ & D-S~$\downarrow$ 
& FAD~$\downarrow$ & IS~$\uparrow$ & KLD~$\downarrow$ \\
\midrule

\multicolumn{7}{l}{{\color[HTML]{656565} \textit{FoleyBench-Long results.}}} \\
LOVA~\cite{cheng2025lova} & 0.237 & 0.102 & 1.20 & \textbf{26.2} & \textbf{5.02} & 2.44 \\
VTA-LDM~\cite{xu2024video} & 0.147 & 0.091 & 1.22 & 83.2 & 1.27 & \textbf{2.19} \\
MMAudio~\cite{cheng2025mmaudio} & \textbf{0.239} & \textbf{0.174} & \textbf{0.638} & 27.5 & 3.87 & 2.40 \\
\midrule

\multicolumn{7}{l}{{\color[HTML]{656565} \textit{FoleyBench results.}}} \\
LOVA~\cite{cheng2025lova} & 0.209 & 0.167 & 1.15 & 20.7 & 7.61 & 3.15 \\
VTA-LDM~\cite{xu2024video} & 0.221 & 0.138 & 1.21 & 15.7 & 7.27 & 3.13 \\
MMAudio~\cite{cheng2025mmaudio} &\textbf{ 0.306} &\textbf{ 0.331 }& \textbf{0.447} & \textbf{8.76} & \textbf{11.2} & \textbf{2.43} \\
\bottomrule
\end{tabular}
}
\vspace{-0.1in}
\end{table}

\vspace{-0.1 in}
\section{Conclusion}
\label{sec:conclusion}
\vspace{-0.1 in}
We introduce FoleyBench, a benchmark for evaluating V2A generation models on Foley-style clips with precise temporal and semantic grounding. Comprising 5,000 human-validated video–audio–caption triplets, FoleyBench enables comprehensive evaluation of audio quality, alignment, and diversity. Our experiments show that widely used benchmarks often include content unsuitable for Foley evaluation, which can lead to misleading conclusions about model performance. In contrast, FoleyBench surfaces meaningfully different results compared to the popular VGGSound test set, particularly with respect to audio fidelity. Through our analysis, we uncover several insights into current model limitations. We find that models struggle to render high-fidelity discrete sounds, even when they are well synchronised; performance degrades significantly on long-form videos, with top short-clip models like MMAudio suffering a severe drop in audio quality; and text conditioning provides a critical semantic prior that models struggle to derive from video alone. By releasing this benchmark and the accompanying data pipeline, we aim to drive the development of more accurate and visually grounded V2A systems.

\newpage
\footnotesize
\vspace{-0.1 in}
\section{Acknowledgments}
\label{sec:acknowledgments}
\vspace{-0.1 in}
This work used Bridges2 at PSC by allocation SOC240007P from the Advanced Cyberinfrastructure Coordination Ecosystem: Services \& Support (ACCESS) program, which is supported by U.S. National Science Foundation grants \#2138259, \#2138286, \#2138307, \#2137603, and \#2138296. We thank Prem Seetharaman, Uri Nieto, and Justin Salamon (Adobe) for their valuable feedback and for running MultiFoley on our dataset, and Heng Wang (University of Sydney) and Chunghsin Yeh (Dolby) for evaluating MaskVAT and V2A-mapper on our dataset.


\vspace{-0.1 in}
\bibliographystyle{IEEEbib}
\bibliography{strings,refs}

\newpage
\appendix

% -------------------------------------------------------------------------
\normalsize
\vspace{-0.1 in}
\section{Baselines}
\label{sec:baselines}
\vspace{-0.1 in}
\noindent\textbf{V-AURA.} Viertola et al.~\cite{viertola2025temporally} propose an autoregressive model that generates audio by sequentially predicting audio tokens, designed to capture fine-grained temporal alignment using high-frame-rate visual features.

\noindent\textbf{DiffFoley.} Luo et al.~\cite{luo2023diff} introduce a diffusion-based model that synthesizes Foley sounds by iteratively denoising a random signal conditioned on visual features.

\noindent\textbf{VTA-LDM.} Xu et al.~\cite{xu2024video} develop a Latent Diffusion Model (LDM) that generates audio in a compressed latent space, conditioned on both video and text inputs for improved control and efficiency.

\noindent\textbf{FoleyCrafter.} Zhang et al.~\cite{zhang2024foleycrafter} present a diffusion model that incorporates both global video context and fine-grained frame-level features to generate highly synchronized audio.

\noindent\textbf{LOVA.} Cheng et al.~\cite{cheng2025lova} introduce a diffusion model specifically designed for long-form video-to-audio generation, capable of producing temporally coherent audio for extended video clips.

\noindent\textbf{V2A-Mapper.} Wang et al.~\cite{wang2024v2a} introduce a lightweight mapping method that projects video embeddings into the latent space of a pretrained audio generator, enabling modular video-to-audio synthesis.

\noindent\textbf{CAFA.} Benita et al.~\cite{benita2025cafa} propose an adapter-based approach that injects visual guidance into a pretrained text-to-audio model using a ControlNet-like architecture, enabling efficient V2A generation.

\noindent\textbf{SpecMaskFoley.} Zhong et al.~\cite{zhong2025specmaskfoley} present a transformer-based model that adapts the masked prediction framework by iteratively filling in masked portions of an audio spectrogram conditioned on visual inputs.

\noindent\textbf{Seeing \& Hearing.} Xing et al.~\cite{xing2024seeing} propose a two-stage model that first generates a descriptive text caption from the video and then uses a text-to-audio model to synthesize the corresponding sound.

\noindent\textbf{MaskVAT.} Pascual et al.~\cite{pascual2024masked} develop a transformer-based model using a masked prediction objective to generate audio spectrograms from visual inputs, adapting the MaskGIT framework for the audio domain.

\noindent\textbf{MultiFoley.} Chen et al.~\cite{chen2025video} introduce a diffusion transformer-based model designed for multimodal control (text, audio, and video). It is trained on a combination of internet videos and professional, high-quality SFX libraries to produce full-bandwidth (48kHz) audio.

\noindent\textbf{MMAudio.} Cheng et al.~\cite{cheng2025mmaudio} introduce a large-scale, jointly trained model using a flow-matching architecture to generate high-fidelity audio conditioned on both video and text inputs.
\vspace{-0.1 in}
\section{Prompts}
\label{sec:prompts}
\vspace{-0.1 in}

The FoleyBench curation pipeline uses two sequential prompts to automatically filter videos and enrich them with metadata.

The first prompt (Figure~\ref{fig:hypothesis_prompt}), given to Gemini 2.5 Pro~\cite{GoogleDeepMind2025Gemini2_5}, performs the primary filtration step. It issues an Accept/Reject verdict, accepting a video only if all sounds are visually and causally grounded while strictly filtering out any speech or music. The model also generates a one-line caption and categorizes the clip by its sound envelope (Discrete/Continuous), source complexity (Single-source/Multi-source), and acoustic focus (Background/Action), returning all information in a structured JSON.

Next, a second prompt (Figure~\ref{fig:cat_prompt}) performs taxonomic labeling on accepted videos. This prompt operates on text-only inputs—the previously generated caption and a coarse YAMNet label—to avoid the cost of reprocessing the video. It maps the clip to the most appropriate categories from AudioSet and the Universal Category System (UCS), using the full taxonomies as context to ensure it outputs valid labels and a rationale.

\begin{figure*}[htbp]
\centering
\begin{tcolorbox}[colback=gray!10, colframe=gray!50, boxrule=0.5pt, arc=3pt, left=6pt, right=6pt, top=6pt, bottom=6pt]
\small
\begin{verbatim}
video_prompt = """
You are given a short video clip from a large video dataset. You are evaluating this video 
for inclusion in a benchmark that tests models on video-to-audio generation. Your goal is 
to accept the video if it is possible to predict the audio from the video alone. Any sound 
whose **timing** and **type** cannot be inferred from the video for any frame must be
rejected.
Your task is to:
1. Make a binary decision: **Accept** or **Reject** this video for inclusion in the
benchmark based on if the audio can be predicted from the video alone for any frame. More
specifically:
   - Accept the video if:
     - All sounds are visually grounded and causally linked to visible events in the video. 
     It should be possible to predict the audio (both the content and the timing) from the
     video alone.
     - There is **no speech** in any frame. Vocalizations (e.g., crying, laughing) are
     acceptable only if they are clearly caused by visible actions. Standard **crowd
     background noise** is acceptable, but **shouting, screaming, cheering something** is
     **not allowed**.
     - There is **no music**. Even faint background music is not allowed. 
   - Reject the video if:
     - Any sound is not clearly caused by or implied by visible actions. It is not possible 
     to predict the audio (content, timing, or both) from the video alone.
     - The video has no visible activity like a static image or the video is silent.
     - There is any **speech or music**, even faintly, or in the background.
2. Categorize the video according to the following:
   a. **Discrete vs Rest**: If the sounds in this video are distinct, impulse-like events
   (e.g. hammer banging, footsteps, tennis ball bounce, door slamming) in an otherwise
   silent clip, label as "Discrete", label all others as "Rest".
   b. **Single-source vs Multi-source**: Are the sounds in the video from a single source
   (e.g. person clapping, dog barking), or are there multiple sound sources in the clip
   (e.g. people talking and footsteps, thunder and car engine)? Label as "Single-source"
   or "Multi-source".
   c. **Background vs Action vs Combined**: Classify whether the sound is (i) mostly
   background/environmental noise (e.g., wind, rain, city ambience), (ii) mostly
   foreground action sounds (e.g., hitting, bouncing, clapping), or (iii) a combination of
   both (e.g., someone hitting a ball during a thunderstorm). Label as "Background",
   "Action", or "Combined".
3. Write a **single-line caption** that summarizes the sounds based on the main visible
activity in the video.
4. Provide a brief explanation (1 sentence) for why you made the classification and
verdict decision. Include reasoning about what was seen in the video and how it led to 
your categorization.

Return your output in the following structured JSON format:
```json
{
  "caption": "A short, descriptive caption of the visible content",
  "discrete_vs_rest": "Discrete" or "Rest",
  "source_type": "Single-source" or "Multi-source",
  "sound_type": "Background" or "Action" or "Combined",
  "verdict": "Accept" or "Reject",
  "reason": "Your explanation for the above choices"
}
```
"""
\end{verbatim}
\end{tcolorbox}
\caption{Prompt for video content filtration and categorization.}
\label{fig:hypothesis_prompt}
\end{figure*}

\begin{figure*}[htbp]
\centering
\begin{tcolorbox}[colback=gray!10, colframe=gray!50, boxrule=0.5pt, arc=3pt, left=6pt, right=6pt, top=6pt, bottom=6pt]
\small
\begin{verbatim}
category_prompt = """
You are an expert audio taxonomy annotator. Given a VIDEO_CAPTION and a NOISY_LABEL (a
coarse label from a sound classifier), map the video to:
1) Exactly ONE top-level UCS category key from the provided UCS_categories (keys only), and
2) Exactly ONE top-level AudioSet category key from the provided audioset_categories 
(keys only).

Return ONLY valid JSON with this schema (no extra text, no code fences):
{{
  "ucs": {{
    "category": "<UCS key>",
    "rationale": "<short explanation>"
  }},
  "audioset": {{
    "category": "<AudioSet key>",
    "rationale": "<short explanation>"
  }}
}}

UCS_categories:
{json.dumps(UCS_categories, ensure_ascii=False, indent=2)}

AudioSet_categories:
{json.dumps(audioset_categories, ensure_ascii=False, indent=2)}
"""
\end{verbatim}
\end{tcolorbox}
\caption{Prompt for video content categorization into AudioSet and UCS categories.}
\label{fig:cat_prompt}
\end{figure*}

\vspace{-0.1 in}
\section{Dataset Diversity}
\label{sec:diversity}
\begin{figure*}[ht]
\centering
\includegraphics[width=1\linewidth]{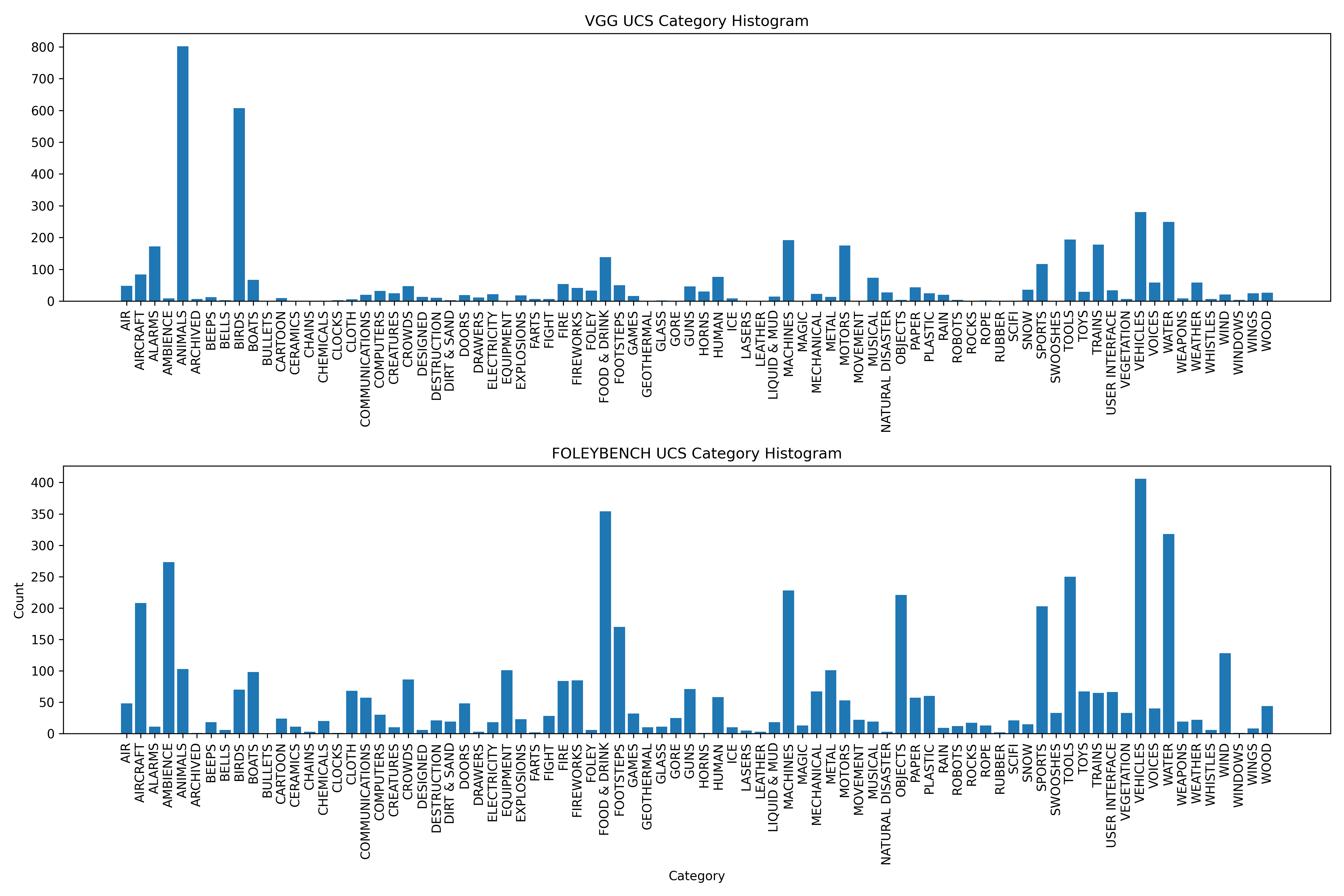}
\caption{\textbf{Comparison of UCS Category Distributions.} The top histogram shows the distribution of the filtered VGGSound test set (after removing music and speech videos), which is heavily skewed towards categories like ANIMALS and BEEPS. The bottom histogram shows the more balanced and uniform distribution of FoleyBench, which provides better coverage across a wider range of Foley-relevant sound categories.}
\label{fig:ucs_dist}
\end{figure*}

\vspace{-0.1 in}
Figure~\ref{fig:ucs_dist} compares the category diversity of FoleyBench against the subset of the VGGSound test set remaining after being filtered for speech and music. The comparison highlights that even after this initial filtering, the VGGSound data remains poorly suited for comprehensive evaluation due to a highly imbalanced distribution. Its content is heavily skewed and dominated by a few common sound types (such as ANIMALS and BIRDS), leaving many other Foley-relevant classes under-represented.

In contrast, FoleyBench provides a visibly more uniform and balanced distribution across the UCS taxonomy, ensuring broader coverage. This visual observation is confirmed by the quantitative analysis in Section \ref{subsec:comparison}: FoleyBench achieves a higher Shannon entropy (5.35 vs. 4.73), indicating greater overall diversity. It also better mitigates data sparsity, with only 13.4\% of its categories containing three or fewer videos, compared to 24.3\% in the filtered VGGSound set. This balanced distribution ensures evaluations are not biased by a model's performance on a handful of common sounds, offering a more reliable benchmark for general Foley synthesis.

% \section{Video Examples}
% \label{sec:videeo_examples}

% Figure~\ref{fig:examples} illustrates representative clips from FoleyBench. Each clip contains a visible sound source and exemplifies the diversity of audio events, including discrete impacts, continuous actions, single-source and multi-source scenes, and combinations of foreground and background sounds. These examples highlight the richness of the dataset and its suitability for evaluating video-to-audio generation models.

% \begin{figure*}
%     \centering
%     \includegraphics[width=0.8\linewidth]{figures/video_examples.png}
%     \caption{Example videos from FoleyBench}
%     \label{fig:examples}
% \end{figure*}

% \begin{figure*}
%     \centering
%     \includegraphics[width=1\linewidth]{figures/ucs_histogram_combined.png}
%     \caption{Distribution of VGGSound test-set (after removing music and speech videos) and FoleyBench across UCS categories}
%     \label{fig:histogram}
% \end{figure*}

\end{document}